\documentclass{emulateapj}
\usepackage[squaren, Gray, cdot]{SIunits}

\slugcomment{Accepted for publication in The Astrophysical Journal}

\shorttitle{Subsurface characterization of 67P/Churyumov-Gerasimenko's Abydos Site}
\shortauthors{Brugger et al.}

\begin{document}

\title{Subsurface characterization of 67P/Churyumov-Gerasimenko's Abydos Site}

\author{
B. Brugger\altaffilmark{1},
O. Mousis\altaffilmark{1},
A. Morse\altaffilmark{2},
U. Marboeuf\altaffilmark{3},
L. Jorda\altaffilmark{1},
A. Guilbert-Lepoutre\altaffilmark{4},
D. Andrews\altaffilmark{2},
S. Barber\altaffilmark{2},
P. Lamy\altaffilmark{1},
A. Luspay-Kuti\altaffilmark{5},
K. Mandt\altaffilmark{5},
G. Morgan\altaffilmark{2},
S. Sheridan\altaffilmark{2},
P. Vernazza\altaffilmark{1},
and I.P. Wright\altaffilmark{2}
}

\altaffiltext{1}{Aix Marseille Universit{\'e}, CNRS, LAM (Laboratoire d'Astrophysique de Marseille) UMR 7326, 13388, Marseille, France (\email{bastien.brugger@lam.fr})}
\altaffiltext{2}{Planetary and Space Sciences, Department of Physics, The Open University, Walton Hall, Milton Keynes MK7 6AA, UK}
\altaffiltext{3}{Physikalishes Institut, Center for Space and Habitability, University of Bern, Switzerland}
\altaffiltext{4}{Institut UTINAM, UMR 6213 CNRS-Universit{\'e} de Franche-Comt{\'e}, Besan\c con, France}
\altaffiltext{5}{Department of Space Science, Southwest Research Institute, 6220 Culebra Rd., San Antonio, TX 78228, USA}

\begin{abstract}

On November 12, 2014, the ESA/Rosetta descent module Philae landed on the Abydos site of comet 67P/Churyumov-Gerasimenko. Aboard this module, the Ptolemy mass spectrometer measured a CO/CO$_2$ ratio of 0.07 $\pm$ 0.04 which differs substantially from the value obtained in the coma by the Rosetta/ROSINA instrument, suggesting a heterogeneity in the comet nucleus. To understand this difference, we investigated the physico-chemical properties of the Abydos subsurface leading to CO/CO$_2$ ratios close to that observed by Ptolemy at the surface of this region. We used a comet nucleus model that takes into account different water ice phase changes (amorphous ice, crystalline ice and clathrates), as well as diffusion of molecules throughout the pores of the matrix. The input parameters of the model were optimized for the Abydos site and the ROSINA CO/CO$_2$ measured ratio is assumed to correspond to the bulk value in the nucleus. We find that all considered structures of water ice are able to reproduce the Ptolemy observation with a time difference not exceeding $\sim$50 days, i.e. lower than $\sim$2\% on 67P/Churyumov-Gerasimenko's orbital period. The suspected heterogeneity of 67P/Churyumov-Gerasimenko's nucleus is also found possible only if it is constituted of crystalline ices. If the icy phase is made of amorphous ice or clathrates, the difference between Ptolemy and ROSINA's measurements would rather originate from the spatial variations in illumination on the nucleus surface. An eventual new measurement of the CO/CO$_2$ ratio at Abydos by Ptolemy could be decisive to distinguish between the three water ice structures.

\end{abstract}


\section{Introduction}

On November 12, 2014, the ESA/Rosetta descent module Philae landed at the Abydos site on the surface of comet 67P/Churyumov-Gerasimenko (hereafter 67P/C-G). As part of the scientific payload aboard Philae, the Ptolemy mass spectrometer (Wright~et~al. 2007) performed the analysis of several samples from the surface at Agilkia (Wright~et~al. 2015) and atmosphere at Abydos (Morse~et~al. 2015). The main molecules detected by Ptolemy on the Abydos site were H$_2$O, CO and CO$_2$, with a measured CO/CO$_2$ molar ratio of 0.07~$\pm$~0.04. Meanwhile, the CO/CO$_2$ ratio has also been sampled in 67P/C-G's coma between August and September 2014 by the ROSINA Double Focusing Mass Spectrometer (DFMS; Balsiger~et~al. 2007; H{\"a}ssig~et~al. 2013) aboard the Rosetta spacecraft. Strong variations of the CO and CO$_2$ production rates, dominated by the diurnal changes on the comet, have been measured by ROSINA, giving CO/CO$_2$ ratios ranging between 0.50~$\pm$~0.18 and 1.62~$\pm$~1.34 over the August-September 2014 time period (H{\"a}ssig~et~al. 2015). Large fluctuations correlated with the sampled latitudes have also been observed and explained either by seasonal variations  or by a compositional heterogeneity in the nucleus (H{\"a}ssig~et~al. 2015). Further investigation of the coma heterogeneity performed by Luspay-Kuti~et~al. (2015) in the southern hemisphere of 67P/C-G at a later time period led to conclusions in favor of compositional heterogeneity. This latter hypothesis is also reinforced by the Ptolemy measurement of the CO/CO$_2$ ratio at the Abydos site, which is found outside the range covered by the ROSINA measurements (Morse~et~al. 2015).

Here, we aim at investigating the physico-chemical properties of the Abydos subsurface which can reproduce the CO/CO$_2$ ratio observed by Ptolemy, assuming that the composition of the solid phase located beneath the landing site initially corresponds to the value in the coma. To investigate the possibility of a heterogeneous nucleus for 67P/C-G, we have employed a comet nucleus model with i) an updated set of thermodynamic parameters relevant for this comet and ii) an appropriate parameterization of the illumination at the Abydos site. This allows us to mimic the thermal evolution of the subsurface of this location. By searching for the matching conditions between the properties of the Abydos subsurface and the Ptolemy data, we provide several constraints on the structural properties and composition of Philae's landing site in different cases.

\section{The comet nucleus model}

The one-dimensional comet nucleus model used in this work is described in Marboeuf~et~al. (2012). This model considers an initially homogeneous sphere composed of a predefined porous mixture of ices and dust in specified proportions. It describes heat transmission, gas diffusion, sublimation/recondensation of volatiles within the nucleus, water ice phase transition, dust release, and dust mantle formation. The model takes into account different phase changes of water ice, within: amorphous ice, crystalline ice and clathrates. The use of a 1D model is a good approximation for the study of a specific point at the surface of the comet nucleus, here the Abydos landing site. However, since 67P/C-G's shape is far from being a sphere (Sierks~et~al. 2015), we have parameterized the model in a way that correctly reproduces the illumination conditions at Abydos. This has been made possible via the use of the 3D shape model developed by Jorda~et~al. (2014) which gives the coordinates of Abydos on the surface of 67P/C-G's nucleus, as well as the radius corresponding to the Abydos landing site and the normal to the surface at this specific location. The Abydos landing site is located just outside the Hatmethit depression, at the very edge of the area illuminated during the mapping and close observation phases of the Rosetta mission, roughly until December 2014. It is also at the edge of a relatively flat region of the small lobe illuminated throughout the perihelion passage of the comet. Geomorphologically, Abydos is interpreted as being a rough deposit region composed of meter-size boulders (Lucchetti~et~al. 2016). Other geometric parameters specific to 67P/C-G, such as the obliquity and the argument of the subsolar meridian at perihelion, are calculated from the orientation of the spin axis computed during the shape reconstruction process (Jorda~et~al. 2014). Table~1 summarizes the main parameters used in this work. The porosity and dust/ice ratio of the cometary material are set in the range of measurement of 80~$\pm$~5~\% (Kofman~et~al. 2015) and 4~$\pm$~2 (Rotundi~et~al. 2014), respectively. These two parameters are linked through the density of the cometary material, and are set to be compatible with the preliminary value determined by Jorda~et~al. (2014) (510~$\pm$~20~kg/m$^3$). 67P/C-G's thermal inertia is estimated to be in the 10--150~W~K$^{-1}$~m$^{-2}$~s$^{1/2}$ range based on the measurement obtained by the Rosetta/VIRTIS instrument (Leyrat~et~al. 2015). According to the same study, regions surrounding Abydos are characterized by a thermal inertia in the lower half of this range. We have therefore chosen a low thermal inertia close to 50~W~K$^{-1}$~m$^{-2}$~s$^{1/2}$.

In addition to water ice and dust, the solid phase of our model includes CO and CO$_2$ volatiles. Although coma abundances do not necessarily reflect those in the nucleus, they constitute the most relevant constraint available on its composition. We thus considered the CO/CO$_2$ ratio (1.62~$\pm$~1.34) measured by ROSINA on August 7, 2014 at 18 hours, as representative of the bulk ice composition in the nucleus, and more specifically in the Abydos subsurface. This ratio is derived from the CO/H$_2$O and CO$_2$/H$_2$O ROSINA measurements performed at this date, which are equal to 0.13~$\pm$~0.07 and 0.08~$\pm$~0.05, respectively (H{\"a}ssig~et~al. 2015). We selected the date of August 7 because i) the corresponding ROSINA measurements were performed at high northern latitudes where the Abydos site is located, and ii) the large CO/CO$_2$ range obtained at this moment covers all values measured by ROSINA at other dates (including the late value obtained by Le~Roy~et~al. (2015) on October 20 for the northern hemisphere, namely CO/CO$_2$~=~1.08).

The three main phases of ices identified in the literature, namely crystalline ice, amorphous ice and clathrate phase, are considered in this work. Outgassing of volatiles in 67P/C-G could then result from the sublimation of ices, amorphous-to-crystalline ice phase transition, or destabilization of clathrates in the crystalline ice, amorphous ice and clathrates cases, respectively. Because the properties of volatiles trapping in the nucleus matrix strongly depend on the considered icy phase, the following models have been considered:

\paragraph{Crystalline ice model} Water ice is fully crystalline, meaning that no volatile species are trapped in the water ice structure. Here, CO and CO$_2$ are condensed in the pores of the matrix made of water ice and dust;

\paragraph{Amorphous ice model} The matrix itself is made of amorphous water ice with a volatile trapping efficiency not exceeding $\sim$10\%. In this case, the cumulated mole fraction of volatiles is higher than this value, implying that an extra amount of volatiles is crystallized in the pores. With this in mind, we consider different distributions of CO and CO$_2$ in the both phases of this model;

\paragraph{Clathrate model} Water ice is exclusively used to form clathrates. Similarly to amorphous ice, clathrates have a maximum trapping capacity ($\sim$17\%). The extra amount of volatiles, if any, also crystallizes in the pores. In our case, however, CO is fully trapped in clathrates and escapes only when water ice sublimates. In contrast, we assume that solid CO$_2$ exists in the form of crystalline CO$_2$ in the pores of the nucleus because this molecule is expected to condense in this form in the protosolar nebula (Mousis~et~al. 2008).

\begin{deluxetable}{lcl}
\tablecaption{Modeling parameters for the nucleus}
\tablehead{
\colhead{Parameter} & \colhead{Value} & \colhead{Reference}
}
\startdata
Rotation period	(hr)																					& 12.4				& Mottola~et~al. (2014)		\\
Obliquity ($\degree$)																					& 52.25				&							\\
Subsolar meridian $\Phi$ ($\degree$) \footnote{Argument of subsolar meridian at perihelion.}			& -111				&							\\
Co-latitude ($\degree$) \footnote{Angle between the normal to the surface and the equatorial plane.}	& -21				&							\\
Initial radius (km)																						& 2.43				&							\\
Bolometric albedo (\%)																					& 1.5				& Fornasier~et~al. (2015)	\\
Dust/ice mass ratio																						& 4~$\pm$~2			& Rotundi~et~al. (2014)		\\
Porosity (\%)																							& 80~$\pm$~5		& Kofman~et~al. (2015)		\\
Density (kg/m$^3$)																						& 510~$\pm$~20		& Jorda~et~al. (2014)		\\
I (W~K$^{-1}$~m$^{-2}$~s$^{1/2}$) \footnote{Thermal inertia.}											& 50				& Leyrat~et~al. (2015)		\\
CO/CO$_2$ initial ratio																					& 1.62~$\pm$~1.34	& H{\"a}ssig~et~al. (2015)	\\
\enddata
\end{deluxetable}

\section{Thermal evolution of the subsurface at Abydos}

Our results show that the illumination at the surface of the Abydos site is a critical parameter for the evolution of the nucleus, regardless the considered ice structure. Consequently, all three models described in Section~2 present the same behavior up to a given point. We first describe the characteristics displayed in common by the three models by presenting the thermal evolution of the crystalline ice model, before discussing the variations resulting from the different assumptions on the nature of ices. Figure~1 shows the time evolution of the nucleus stratigraphy, which corresponds to the structural differentiation occurring in the subsurface of the Abydos site. This differentiation results from the sublimation of the different ices. After each perihelion passage, the sublimation interfaces of CO and CO$_2$ reach deeper layers beneath the nucleus surface, with a progression of $\sim$20 m per orbit. The CO sublimation interface always propagates deeper than its CO$_2$ counterpart because of the higher volatility of the former molecule. On the other hand, because surface ablation is significant, the progression of these interfaces is stopped by the propagation of the water sublimation front after perihelion. This allows the Abydos region to present a ``fresh'' surface after each perihelion.

\begin{figure}[h]
\begin{center}
\includegraphics[width=9cm]{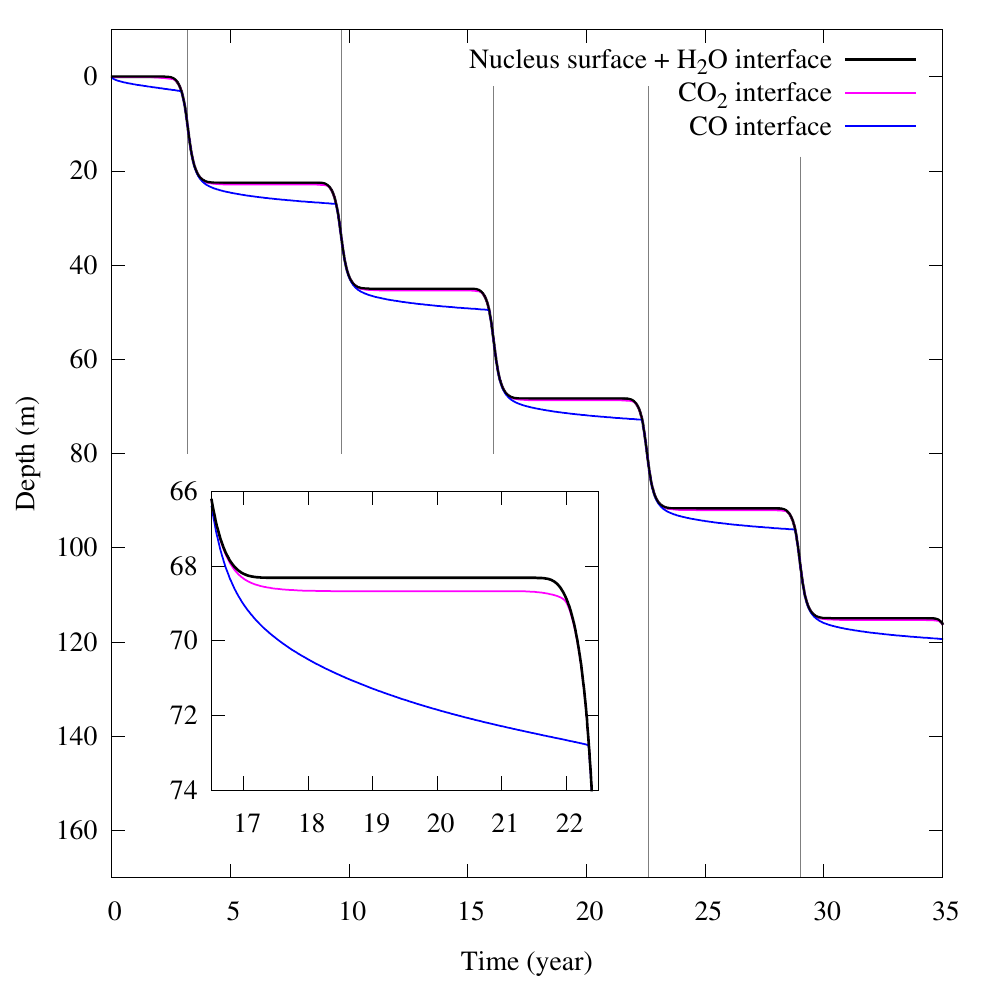}
\caption{Stratigraphy of the Abydos subsurface represented during 35 years of evolution ($\sim$5 orbits of 67P/C-G), showing the interfaces of sublimation of considered volatile species. The surface ablation occurs at each perihelion (represented by the vertical lines) and reaches all interfaces.}
\end{center}
\end{figure}

At the surface of the Abydos site, the outgassing rates of CO and CO$_2$ vary with the illumination conditions, reaching maxima at perihelion and minima at aphelion. Because the sublimation interface of CO$_2$ is closer to the surface, its production rate is more sensitive to the illumination conditions than that of CO. As a result, the outgassing rate of  CO$_2$ presents important variations with illumination while that of CO is less affected. This difference strongly impacts the evolution of the CO/CO$_2$ outgassing ratio at the surface of Abydos (see Figure~2). Close to perihelion, this ratio crosses the range of values measured by Ptolemy (0.07~$\pm$~0.04) and reaches a minimum. Note that the CO/CO$_2$ outgassing ratio presents spikes during a certain period after perihelion. These spikes appear when the CO and CO$_2$ interfaces of sublimation are dragged out to the surface by ablation and result from temperature variations induced by diurnal variations of the insolation.

We define $\Delta t$ as the time difference existing between the Ptolemy CO/CO$_2$ observations (here 0.07 measured on November 12, 2014) and the epoch at which our model reproduces these data (see Figure~2). In each case investigated, we vary the input parameters of the model to minimize the value of $\Delta t$ (see Table~2 for details). We have also defined the quantity $f_t = \Delta t / P_{orb}$, namely the fraction of 67P/C-G's orbital period ($P_{orb} = 6.44$ yr) corresponding to $\Delta t$. The results of our simulations are indicated below.

\begin{figure}[h]
\begin{center}
\includegraphics[width=9cm]{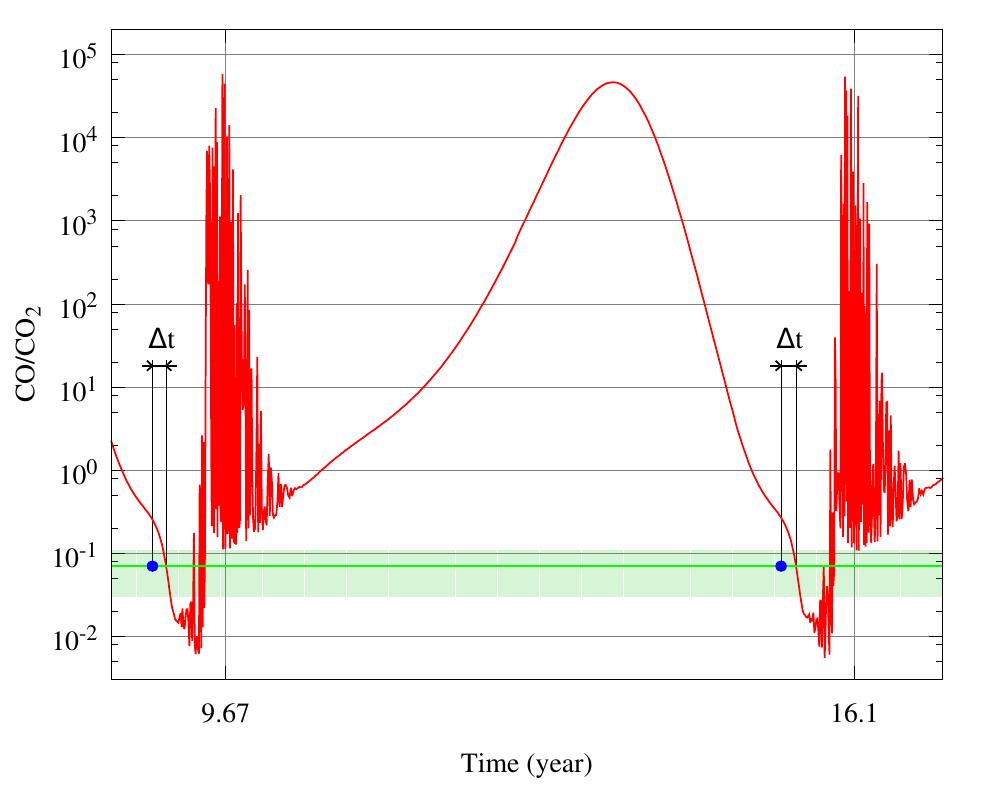}
\caption{Evolution of the CO/CO$_2$ outgassing ratio at Abydos in the case of the crystalline ice model, during one orbit. The green line and green area represent the Ptolemy central value and its range of uncertainty, respectively. The blue dots correspond to the measurement epoch (November 12, 2014). Vertical lines correspond to passages at perihelion. See text for a  description of $\Delta t$.}
\end{center}
\end{figure}

\begin{deluxetable}{lccc}
\tablecaption{Optimized set of parameters for the three models}
\tablehead{
\colhead{} & \colhead{Crystalline} & \colhead{Amorphous} & \colhead{Clathrate}\\ 
\colhead{} & \colhead{ice model} & \colhead{ice model} & \colhead{model}
}
\startdata
Parameters										&					&					&				\\
\hline
Dust/ice mass ratio								& 6					& 4					& 4				\\
Porosity (\%)									& 78				& 76				& 76			\\
Density (kg/m$^3$)								& 516				& 510				& 519			\\
I (W K$^{-1}$ m$^{-2}$ s$^{1/2}$) \footnote{Thermal inertia resulting from the different water ice conductivities.}		& $\sim$60	& 40--60	& $\sim$50	\\
CO/CO$_2$ initial ratio							& 0.46				& 1.62				& 0.46			\\
CO/H$_2$O initial abundance						& 6\%				& 13\% \footnote{2.8\% trapped in amorphous ice, 10.2\% condensed in the pores (Marboeuf~et~al. 2012).}	& 6\%	\\
CO$_2$/H$_2$O initial abundance					& 13\%				& 8\% \footnote{4.2\% trapped in amorphous ice, 3.2\% condensed in the pores (Marboeuf~et~al. 2012).}	& 13\%	\\
\hline
Results											&					&					&				\\
\hline
$\Delta t$ (day)								& 52				& 4					& 34			\\
$f_t$ (\%)										& 2.2				& 0.17				& 1.4			\\
\enddata
\end{deluxetable}

\paragraph{Crystalline ice model} In this case, we find that the initial CO/CO$_2$ ratio and the dust/ice ratio adopted in the nucleus have a strong influence on $\Delta t$. This quantity is minimized i) when the adopted dust/ice ratio becomes higher than those found by Rotundi~et~al. (2014) and ii) if the selected initial CO/CO$_2$ ratio is lower than the nominal value found by ROSINA (H{\"a}ssig~et~al. 2015). Figure~3 shows the evolution of $\Delta t$ as a function of the dust/ice ratio for two different values of the initial CO/CO$_2$ ratio, namely 1.62 (the central value) and 0.46 (close to the lower limit). These results confirm the aforementioned trend: with CO/CO$_2$ = 0.46 and a dust/ice ratio of 6 or higher, the Ptolemy's measurement epoch is matched with $\Delta t$~=~52~$\pm$~27 days or lower, which corresponds to $f_t$ lower than $\sim$2\%. These results can be explained by the thermal conductivity of crystalline water ice, which is in the 3--20~W~m$^{-1}$~K$^{-1}$ range (Klinger 1980), considering the temperatures in the comet. Because dust has a conductivity of 4~W~m$^{-1}$~K$^{-1}$ (Ellsworth~\&~Schubert 1983), the global conductivity decreases with the increase of the dust/ice ratio in the nucleus. Since heating in the nucleus is mostly provided by surface illumination, a low conductivity increases the temperature gradient between the upper and deeper layers (where the CO$_2$ and the CO sublimation interfaces are located, respectively). This gradient enhances more the sublimation rate of CO$_2$ ice than that of CO ice, leading to smaller CO/CO$_2$ outgassing ratios at the surface and to a smaller $\Delta t$ (see Figure~2).

\begin{figure}
\begin{center}
\includegraphics[width=9cm]{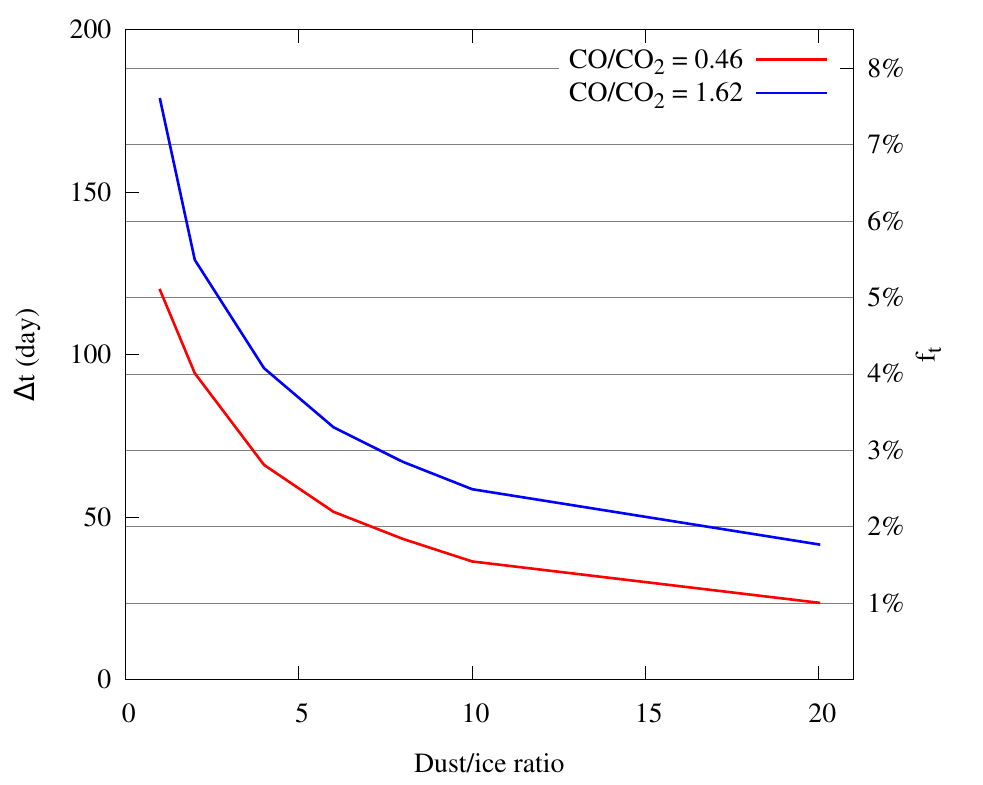}
\caption{Evolution of $\Delta t$ and $f_t$ as a function of the dust/ice mass ratio for two different values of the initial CO/CO$_2$ ratio in the case of the crystalline ice model (see text).}
\end{center}
\end{figure}

\paragraph{Amorphous ice model} Here, values of $\Delta t$ are significantly lower than those obtained with the crystalline ice model; a $\Delta t$ of 4~$\pm$~9 days is obtained using the central values listed in Table~2 for the initial CO/CO$_2$ and dust/ice ratios, leading to $f_t \sim 0.2\%$. Higher values of the CO/CO$_2$ ratio never allow our model matching the Ptolemy data: the CO/CO$_2$ outgassing ratio is increased sufficiently high so that even its minimum becomes higher than the range of measurements performed by Ptolemy. On the other hand, lower CO/CO$_2$ ratios increase $\Delta t$. Interestingly, the results of this model are poorly affected by the considered dust/ice ratio. Within the 0.12--1.35~W~m$^{-1}$~K$^{-1}$ range (Klinger 1980), the conductivity of amorphous water ice lays under those of dust (4~W~m$^{-1}$~K$^{-1}$) and crystalline ices (3--20~W~m$^{-1}$~K$^{-1}$). Since amorphous ice dominates the volatile phase, the mean conductivity is never too far from to that of dust, irrespective of the dust/ice ratio considered within the observed range.

\paragraph{Clathrate model} In this case, low CO/CO$_2$ ratios (still within the range given in Table~1) are required to get values of $\Delta t$ below 50 days (i.e., $f_t$ below 2\%). Similarly to the case of the amorphous ice model, $\Delta t$ is poorly sensitive to the variation of the dust/ice ratio because the conductivity of clathrates (0.5~W~m$^{-1}$~K$^{-1}$; Krivchikov~et~al. 2005a, 2005b) is small compared to that of dust.

\section{Discussion}

Our goal was to investigate the possibility of recovering the value of the CO/CO$_2$ outgassing ratio measured by the Ptolemy instrument at the surface of Abydos. Interestingly, all considered models match the Ptolemy value with $f_t$ lower than $\sim$2\%, provided that an optimized set of parameters is adopted for the Abydos region. Despite the fact it is poorly sensitive to the adopted dust/ice ratio, a nucleus model dominated by amorphous ice (and possibly including a smaller fraction of crystalline ices) gives the best results ($\Delta t$~$\leq$~4~days, i.e. $f_t$~$\leq$~0.2\%) for a primordial CO/CO$_2$ ratio equal to the central value measured by the ROSINA instrument. On the other hand, the crystalline ice and clathrate models require a primordial CO/CO$_2$ ratio close to the lower limit sampled by ROSINA to obtain values of $\Delta t$ under 50 days (i.e. $f_t$ under 2\%). We stress the fact that the CO/CO$_2$ range of validity is used under the assumption that the ROSINA measurements correspond to the bulk nucleus abundances, which is not necessarily true. A second requirement to minimize the value of $\Delta t$ in the crystalline ice model is the necessity to adopt a dust/ice ratio at least equal to or higher than the upper limit determined by Rotundi~et~al. (2014) for 67P/C-G. This is supported by the pictures taken at Abydos by the CIVA instrument (Bibring~et~al. 2007) aboard the Philae module. The very low reflectance of 3--5\% of the Abydos region (Bibring~et~al. 2015) is in agreement with the OSIRIS and VIRTIS reflectance measurements in the visible (Sierks~et~al. 2015) and near-IR (Capaccioni~et~al. 2015), which are consistent with a low ice content in the upper surface layer (Capaccioni~et~al. 2015).

Surface illumination can also greatly influence the CO/CO$_2$ outgassing ratio on 67P/C-G. To quantify this effect, we have simulated a point of 67P/C-G's nucleus which is more illuminated in comparison to Abydos. We have performed a set of simulations at a co-latitude of -52.25$\degree$, a point that receives permanent sunlight around perihelion. At the date when the CO/CO$_2$ outgassing ratio at Abydos is equal to 0.07 (the central value of Ptolemy's range of measurement), we obtain a different value for the CO/CO$_2$ outgassing ratio at this new location, irrespective of the adopted model. For the crystalline ice model, the outgassing ratio at the illuminated site reaches a value of 0.11, which is still within Ptolemy's range of measurement. This implies that a different illumination cannot explain a strong variation of the CO/CO$_2$ outgassing ratio if the nucleus presents a homogeneous composition. In this case, the difference between the Ptolemy and ROSINA measurements is clearly due to a heterogeneity in the nucleus composition. On the other hand, the CO/CO$_2$ outgassing ratio at the illuminated site is equal to 0.74 and 0.76 in the cases of the amorphous ice and clathrate models, respectively. These values are within ROSINA's range of measurements, implying that the difference of illumination is sufficient to explain the difference with the CO/CO$_2$ ratio sampled at Abydos, assuming a homogeneous nucleus.

In summary, all possible water ice structures are able to reproduce the observations made by Ptolemy, assuming that the primordial CO/CO$_2$ ratio is the one inferred by ROSINA. Each case requires a unique set of input parameters taken from the range of values inferred by Rosetta and which describes the structure and composition of the material. According to our simulations, a heterogeneity in the composition of 67P/C-G's nucleus is possible only if the nucleus is composed of crystalline ices. However, if we consider different ice phases like amorphous ice or clathrates, the difference between the Ptolemy and ROSINA measurements could simply originate from the variation of illumination between different regions of the nucleus.

In the upcoming months, the Philae module could awaken and allow the Ptolemy mass spectrometer to perform additional measurements of the CO/CO$_2$ ratio. By comparing these new values with the different CO/CO$_2$ outgassing ratios predicted by our three models at the same date, we would be able to see which model is the most reliable, and thus to determine which water ice structure is dominant at the surface of 67P/C-G's nucleus.

\acknowledgements
O.M. acknowledges support from CNES. This work has been partly carried out thanks to the support of the A*MIDEX project (n\textsuperscript{o} ANR-11-IDEX-0001-02) funded by the ``Investissements d'Avenir'' French Government program, managed by the French National Research Agency (ANR). Funding and operation of the Ptolemy instrument was provided by the Science and Technology Facilities Council (Consolidated Grant ST/L000776/1) and UK Space Agency (Post-launch support ST/K001973/1). A.L.-K. acknowledges support from the NASA Jet Propulsion Laboratory (subcontract No. 1496541).

\end{document}